\documentclass[twocolumn,showpacs,preprintnumbers,prb,fleqn]{revtex4}
\usepackage{epsfig}
\usepackage{graphicx}
\usepackage{amsfonts}
\newcommand{\ig}{\includegraphics}
\newcommand{\ct}{\cite}
\newcommand{\bi}{\bibitem}
\newcommand{\be}{\begin{equation}}
\newcommand{\ee}{\end{equation}}
\newcommand{\ba}{\begin{eqnarray}}
\newcommand{\ea}{\end{eqnarray}}

\newcommand{\ket}[1]{|#1\rangle}

\begin{document}
                                                                                
%%%%%%%%%%%%%%%%%%%%%%%%%%%%%%%%%%%%%%%%%%%%%%%%%%%%%%%%%%%%%%%%%%%%%%%%%

\title{ The effect of the three-spin interaction and the next-nearest neighbor
interaction on the quenching dynamics of a transverse Ising model}
\author{Uma Divakaran}
\email{udiva@iitk.ac.in}
\author{Amit Dutta}
\email{dutta@iitk.ac.in}
\affiliation{Department of Physics, Indian  Institute of Technology Kanpur,
Kanpur 208016, India.}
\date{\today}
\begin{abstract}
We study the zero temperature quenching dynamics of various extensions
of the transverse Ising model (TIM) when the transverse field is 
linearly quenched from 
$-\infty$ to $+\infty$ (or zero) at a finite and uniform rate. The rate of
quenching is dictated by a 
characteristic scale given by $\tau$. The density of kinks produced 
in these extended models
while crossing the quantum critical points during the quenching process
 is calculated using a many body generalization of the Landau-Zener transition 
theory. The density of kinks in the final state is found to 
decay as $\tau^{-1/2}$.
In the first model considered here, the transverse Ising Hamiltonian includes an
additional ferromagnetic three spin interaction term of strength $J_3$.
For $J_3<0.5$, the kink density is found to increase monotonically with $J_3$ 
whereas it decreases
with $J_3$ for $J_3>0.5$. The point  $J_3=0.5$ and the transverse field $h=-0.5$
is multicritical where
the density shows a slower decay given by $\tau^{-1/6}$.
We also study the effect of ferromagnetic or antiferromagnetic 
next nearest neighbor (NNN) interactions  on the dynamics of TIM 
under the same quenching scheme.
In a mean field approximation, the transverse Ising Hamiltonians with NNN 
interactions
are identical to the three spin Hamiltonian. The NNN interactions
non-trivially modifies the dynamical behavior, for example
an antiferromagnetic NNN interactions results to  a larger number of kinks
in the final state in comparison to the case when the NNN interaction is
ferromagnetic.

\end{abstract}
\maketitle
\section{Introduction}
The critical  dynamics of  classical systems
have been studied extensively in last three decades
while the study of the  dynamics of a quantum system 
when swept across  a quantum
critical point (QCP) is fairly recent and
not yet fully understood. 
The vanishing of energy gap between the 
ground state and the first excited state of the quantum Hamiltonian 
signals the existence of a QCP\ct{sachdev99,dutta96}. 
At a QCP, the correlation length 
as well as the  relaxation time diverge, a phenomenon known as
 the critical slowing down. 
This diverging timescale makes it impossible for any system
to cross the quantum critical point without excitations from the ground
state. The dynamics therefore  is  non-adiabatic in contrast to
an  adiabatic evolution where the system sticks to the instantaneous 
ground state through out the quenching process. 
In recent years, there has been an upsurge in the study of dynamics
close to a  quantum  critical point
clearly indicating a growing interest in the field\ct{sengupta04,zurek05,
dziarmaga05,damski05,calabrese05,levitov06,polkovnikov05,das06,cincio07,
cramer07,victor07}.

One of such attempts was to extend the Kibble's theory of 
defect production introduced  to explain early universe behavior 
\ct{kibble76} to the 
second order 
quantum phase transitions. This method of calculating the density of
defects is  known as 
the Kibble-Zurek mechanism (KZM)\ct{zurek85}. The theory of KZM for
a classical second order phase transition is based on the
universality of the critical slowing down and leads to the prediction that the
linear dimension  of the  ordered domains scales with the transition time $\tau$ as $\tau^w$ where
$w$ is some combination of critical exponents. 
KZM has been confirmed by 
numerical simulations of time dependent Ginzburg-Landau model \ct {laguna97} 
and also for various experimental systems \ct{ruutu96}. 
On the other hand, for  He-4 superfluid transition, 
KZM could not be verified experimentally\ct{dodd98}.
Hence more experiments are clearly needed to put KZM theory on a stronger 
footing. The same idea has been 
applied to study the dynamics across 
a zero temperature QCP by
different groups\ct{zurek05,dziarmaga05,damski05,polkovnikov05,levitov06,
cincio07,cramer07,victor07}. 

The extended KZM for the zero-temperature quantum transitions relies
on the fact that during the evolution when the system is close to the static
critical point, the relaxation time diverges in a power-law fashion. The
non-adiabatic effects become prominent when the time scale associated
with the  change of the Hamiltonian is of the order of the relaxation time.
The loss of adiabaticity while crossing a quantum critical point can be 
quantified by estimating either the density of defects (e.g., the density
of oppositely oriented spins in Ising models)  in the final state 
\ct{zurek05,dziarmaga05,damski05,levitov06,polkovnikov05} 
or the fidelity of the final state with respect to the 
ground state\ct{zurek05} or the residual energy\ct{kadowaki98,arnab05,
sei05,brooke99,
santoro02,santoro06}. 
The argument given above immediately leads to a $(1/{\sqrt \tau})$-dependence
of the density of defects on the characteristic timescale $\tau$ of the
quenching. 
The residual energy is defined as the
difference between the energy of the evolved ground state and the true  ground
state.
This residual energy for the integrable disorder free systems is 
trivially proportional to the density of kinks with the 
proportionality constant being equal to the strength of interaction. In an
optimization approach popularly known as the ``quantum annealing" 
\ct{kadowaki98,arnab05,sei05,brooke99,santoro02,santoro06}, the
strength of the quantum fluctuations is slowly reduced to zero so that
a disordered and frustrated system of  finite size   
is expected to reach adiabatically its true  classical ground state. In the present literature, the expressions
``annealing" and ``quenching" are used synonymously.
The residual
energy turns out to be a more appropriate measure of non-adiabaticity for
the annealing approach.
In a recent work by Caneva $et.~al.$\ct{caneva07},  
it has been shown that for a disordered quantum Ising spin chain,
the residual energy and the density of kinks 
show different scaling behavior with $\tau$.
Recently a general analysis has been carried out of the breakdown of
the adiabatic limit
in low-dimensional gapless systems \ct{polkovnikov07}.

In this paper, we will concentrate on the estimation of density of defects 
produced during the dynamics of three different types of model Hamiltonians,
all of them being exactly solved, at least in a mean field level, via 
the Jordan Wigner transformation.
These three Hamiltonians are extensions of the TIM with an additional 
interaction term in each and our aim is to study the effect of such 
interactions on the density of defects produced during the quenching.
The additional terms are
i) a ferromagnetic three spin interaction \ct{kopp05}
ii) an antiferromagnetic next nearest 
neighbor interaction and iii) a ferromagnetic next nearest neighbor interaction,
respectively.
We consider the unitary evolution of the system
prepared in the ground state of the initial Hamiltonian which crosses its
equilibrium critical line as the system evolves. As described later, in
all the cases,  the
fermionization of the Hamiltonian reduces it to a quadratic form and hence 
one can reduce the dynamics of a many-body Hamiltonian effectively
 to a $2 \times 2$ Landau Zener problem
\ct{landau} in the fourier representation.  

The paper is organized as follows. Section II includes a detailed 
discussion on the analytical diagonalization of the 
transverse Ising Hamiltonian 
with a three spin interaction term. In section III, we have described the
transverse quenching scheme along with the results for the above model.
We have presented a comparison between the three spin Hamiltonian and 
the Hamiltonians with next nearest neighbor interactions when treated 
at a mean field level in section IV.
A brief summary of the work is presented in the concluding section
with a brief discussion based on the recent developments in this field.

\section{Model and the Phase diagram}
The Hamiltonian of a one-dimensional three spin interacting transverse Ising 
system is given by\ct{kopp05}
\ba
H=&-\frac 1 {2} \{&\sum_{i}\sigma^{z}_{i}[h+J_3\sigma^{x}_{i-1}\sigma^{x}_{i+1}]\nonumber\\
&+&J_x\sum\sigma^{x}_{i}\sigma^{x}_{i+1}\},
\ea
where $\sigma^z$ and $\sigma^x$ are non-commuting Pauli spin
matrices, $J_x$ is the strength of the nearest neighbor
ferromagnetic interaction while $J_3$ denotes the strength
of the three spin interaction. In the limit $J_3\to 0$, the model
reduces to the celebrated transverse Ising model studied extensively in
recent years\ct{sachdev99,dutta96}. By a duality transformation\ct{kogut79},
the above Hamiltonian
can be mapped to a transverse $XY$ model with competing (ferro-antiferro) 
interactions in the $x$ and $y$ components of the spin\ct{kopp05}.
Interestingly, even in the presence of the three spin interaction term, 
the Hamiltonian given by Eq.~(1) is exactly solved by the Jordan-Wigner (JW)
transformation\ct{lieb61,pfeuty70,kogut79} which maps this interacting
spin system to a system of noninteracting spinless fermions. 
Moreover, this three spin term is 
found to be irrelevant in determining the quantum critical behavior of the
system. The  critical exponents  are the same as that of Ising model in a
transverse field except for the case $J_x=0$ ,and $J_3=h$. For the
sake of completeness, let us now provide a brief discussion on the 
diagonalization of the Hamiltonian.

In the JW-transformation, the Pauli matrices
are transformed to a set of fermionic operators ($c_i$) defined
as 
\ba
c_i&=&\sigma_i^{-} \exp(-i\pi\sum_{j=1}^{i-1}\sigma_{j}^{\dagger}\sigma_{j}^-)
\nonumber\\
\sigma^z_i&=&2c_i^{\dagger}c_i-1
\ea
with $\sigma^{\dagger}=(\sigma^x+\rm i\sigma^y)/2$ and $\sigma^-=(\sigma^x-\rm i\sigma^y)/2$, and satisfy the standard anticommutation relations
$$\{c_i^{\dagger},c_j\}=\delta_{ij},~~\{c_i^{\dagger},c_j^{\dagger}\}=
\{c_i,c_j\}=0.$$ 
We shall work in the basis in which $\sigma^z$ is diagonal so that
the presence of a fermion at a particular site $i$ corresponds to
an up spin (i.e., eigenvalue $+1$ of the operator $\sigma_i^z$) at that site.
 Using a periodic boundary condition, the Fourier transform of the
Hamiltonian can be cast in the form 
\ba
H=-[\sum_{k>0} (h+{\rm cos}k-J_3{\rm cos}2k)(c_k^{\dagger}c_k+c_{-k}^{\dagger}
c_{-k})\nonumber\\
+{\rm i}({\rm sin}k-J_3{\rm sin}2k)(c_k^{\dagger}c_{-k}^{\dagger}+c_kc_{-k})].
\ea
Clearly, in the momentum representation of c-fermions, 
the Hamiltonian is quadratic and is translationally invariant.
Using the Bogoliubov transformation, the Hamiltonian can be diagonalized
to the form $-\sum_k\epsilon_k\eta_k^{\dagger}\eta_k$ where $\eta_k$ are
the Bogoliubov quasiparticles and $\epsilon_k$
is the excitation energy or gap given by \ct{lieb61,kopp05}
\ba
\epsilon_{k}=(h^2+1+J_3^2+2h{\rm cos}k-2hJ_3{\rm cos}2k-2J_3{\rm cos}k)^{1/2}
\ea
with $J_x$ set equal to unity.

It can be easily shown that the gap of the spectrum vanishes at $h=J_3+1$ and 
also at $h=J_3-1$  with ordering (or mode-softening)
 wave vectors $\pi$ and $0$ respectively.
These two lines correspond to quantum phase transitions from 
a ferromagnetically ordered phase to a quantum paramagnetic phase 
with the associated exponents
being the same as the transverse Ising model \ct{pfeuty70}. The wave vector at
which the minima of 
$\epsilon_k$ (Eq.~4) occurs, gets shifted from $k=0$ to $k=\pi$ wave vector 
when one crosses the line
$h=J_3$.
Moreover, there
is an additional phase transition at $h=-J_3$.
This transition belongs to the universality
class of the anisotropic transition observed in the transverse XY-model 
dual to the
Hamiltonian (1)\ct{bunder99}
and the phase boundary is flanked by the incommensurate phases on either side 
with ordering wave vector given by
\be
{\rm cos}k=\frac{h-J_3}{4hJ_3}~~.
\ee
This  incommensurate wave vector picks up a value $k_o$ such that
$\cos k_0=1/2J_3$
at the phase boundary. 
Obviously, for $J_3<0.5$, the anisotropic phase transition
can not occur. The equilibrium phase diagram of the model is shown in Fig.~1.
\begin{figure}[h]
\ig[height=2.2in]{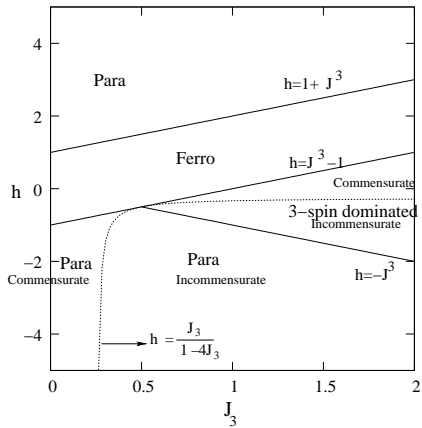}
\caption{Equilibrium phase diagram of the three spin interacting Ising model.
Solid lines show phase boundaries and dotted line marks the boundary
between the incommensurate and the commensurate phase.} 
\end{figure}

\section{Transverse Quenching and Results}

The dynamics of the three spin interacting TIM is found to be very 
interesting due to the fact that the system crosses various quantum
critical lines during the process of dynamics.
As mentioned already, the system deviates from the adiabatic evolution
in the neighborhood of a quantum critical point where nonadiabaticity 
dominates due to the divergence of relaxation time.
We shall introduce the time dependence in the Hamiltonian through the 
transverse field  which is linearly quenched from 
$-\infty$ to $+\infty$ at a steady finite rate given by
$h(t)\sim t/\tau$,
where the quenching time $\tau$ determines the rate of quenching
\ct{zurek05,dziarmaga05,damski05}. At time $t=-\infty$ the transverse field
 $h=-\infty$ and hence
all the spins are pointing in the negative $z-$direction.
By virtue of the duality transformation, the 
 transverse quenching of the 3 spin Hamiltonian corresponds to the  anisotropic 
quenching of the transverse  $XY$ model where the interaction term of 
the later Hamiltonian is adiabatically 
changed from $-\infty$ to $\infty$ \ct{victor07}.

Let us recall the Hamiltonian given in Eq.~3 with a time dependent
transverse field $h(t)$. This Hamiltonian can be
split into a sum of independent terms, $H(t)=[\sum_{k>0}H_k(t)]$ where
each $H_k(t)$ operates on a four dimensional Hilbert space spanned by the basis 
vectors $\ket 0,~\ket {k,-k},~\ket k$ and $\ket{-k}$. The vacuum state where no c-particle
is present, is denoted by $\ket 0$ which corresponds to a spin configuration
with all spins pointing in the $-z$-direction. 
The form of the Hamiltonian readily suggests that the parity 
(even or odd) of total number of fermions (given by 
$n_k=c_k^{\dagger}c_k+c_{-k}^{\dagger}c_{-k}$) for each mode is conserved. 
Therefore,
to study the quenching dynamics, it is convenient to project the Hamiltonian 
$H_k(t)$ in the subspace spanned by $\ket 0$ and $\ket {k,-k}$ . The projected
Hamiltonian has a form
\[\left[\begin{array}{ll}
h(t)+\cos k-J_3\cos 2k& i (\sin k-J_3\sin 2k)\\
-i(\sin k-J_3\sin 2k)&-(h(t)+\cos k-J_3\cos 2k)\end{array}\right]\]

In the reduced Hilbert space, any general state can be represented as a 
superposition of $\ket 0$ and $\ket {k,-k}$ 
with time
dependent amplitudes $u_k(t)$ and $v_k(t)$ such that
$\psi_k(t)=u_k(t)\ket 0+v_k(t)\ket {k,-k}$.
The time evolution of the state is given by the Schroedinger equation 
\be
i\partial_t\psi_k(t)=H_k(t)\psi_k(t).
\ee
We shall here  use the initial conditions $u_k(-\infty)=1$ and $v_k(-\infty)=0$
which in the spin language corresponds to the state with  all spins down.
The off diagonal
term $\Delta=\sin k-J_3\sin 2k$ represents the interaction between the two time 
dependent levels with energies
$E_{1,2}=\pm [h(t)+\cos (k)-J_3\cos 2k]$. The zeroes of the off-diagonal term 
yield the mode softening wave vectors $k=0,\pi$ and $\cos^{-1}1/(2J_3)$ 
(provided $J_3>0.5$) at which the system 
becomes quantum critical for appropriate parameter values. At these 
parameter values and wave vectors, the system undergoes a nonadiabatic
transition from its instantaneous ground state. 
A measure of  nonadiabaticity can be obtained by comparing the two level 
problem to the corresponding Landau-Zener transition equations\ct{damski05,levitov06}.
For a completely  adiabatic transition, we expect the final state to be 
described by the  probability amplitudes $u_k(+\infty)=0$ and $v_k(+\infty)=1$,
i.e.,  the complete spin-flip from down to up occurs.
The nonadiabatic transition probability 
$p_k$ is directly given by $|u_k(+\infty)|^2$ where
the probability amplitudes $u_k(t)$ and $v_k(t)$ are normalized
at each instant of time. Equivalently, $p_k$ also  measures the 
probability that the system remains in its initial
state $\ket 0$ even at the final time.
Using the results of Landau-Zener transitions \ct{landau,sei05}, $p_k$ 
is found to be
\be
p_k=|u_k(+\infty)|^2=\exp(-2\pi\gamma)~~{\rm where}~~
\gamma=\frac{\Delta^2}{\frac{d}{dt}(E_1-E_2)}.
\ee
Therefore, in this model 
\be
p_k=\exp[-\pi\tau(\sin k-J_3\sin 2k)^2].
\ee
The variation of $p_k$ as a function of $k$ for different values
of quenching time $\tau$ is shown in Fig.~2. It is to be noted that for 
$J_3<0.5$, there are  peaks at  $-\pi$,0 and $\pi$ in the whole range of
wave vectors from $-\pi$ to $\pi$ whereas for $J_3 > 0.5$ there are additional 
peaks at the incommensurate values $\pm \cos^{-1}(1/2J_3)$. 

\begin{figure}[h]
\ig[height=1.9in]{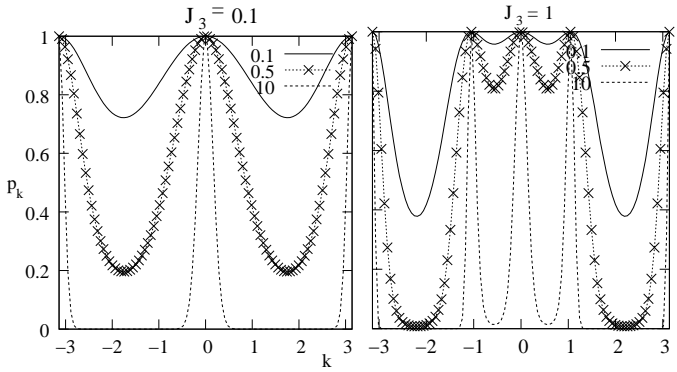}
\caption{Non-adiabatic transition probability $p_k$ for the three 
spin interacting 
Hamiltonian with $J_3=0.1$ in Fig.~2(a) and $J_3=1$ in Fig.~2(b) 
for various $\tau$. It
should be noted that for $J_3=1$, the system undergoes a non-adiabatic
transition at an incommensurate wave vector $k=\pi/3$ and therefore,
there is an additional peak at this wave vector . For large $\tau$, 
$p_k$ is nonzero only for wave vectors very close to the critical modes.
On the other hand, for small values of $\tau$, levels cross quickly resulting
to a non-zero value of $p_k$ for all values of k.}
\end{figure}
As mentioned already, the degree of nonadiabaticity can be quantified 
through the density of kinks $n$
generated at $t=+\infty$ which is obtained by integrating the probability
$p_k$ over the entire range of wave vector.
\be
n=\sum_k p_k=\frac{1}{2\pi}\int_{-\pi}^{\pi}dk~p_k.
\ee
A close inspection of Eq.~8 (see also Fig.~2) shows
that for sufficiently slow quenching (i.e., large $\tau$), only modes
close to the critical modes are excited. One can therefore, to the lowest 
order in k, replace $\sin k$ by $k$ in the exponential of Eq.~8 and arrive
at an approximate analytical expression for density of kinks in the large
$\tau$ limit, given by:

$$ n=\frac{1}{2\pi(1-2J_3)\sqrt \tau}+\frac{1}{2\pi(1+2J_3)\sqrt \tau}~~~
({\rm for}~ J_3<0.5) \eqno(10a)$$
$$n=\frac{2}{2\pi(2J_3-1)\sqrt \tau}+\frac{2}{2\pi(1+2J_3)\sqrt \tau}~~~
({\rm for}~ J_3>0.5).\eqno(10b)$$

%\ba
%n=\frac{1}{2\pi(1-2J_3)\sqrt \tau}+\frac{1}{2\pi(1+2J_3)\sqrt \tau}~~~
%({\rm for}~ J_3<0.5)\nonumber\\
%n=\frac{2}{2\pi(2J_3-1)\sqrt \tau}+\frac{2}{2\pi(1+2J_3)\sqrt \tau}~~~
%({\rm for}~ J_3>0.5) 
%\ea

In Eq.~10(a), the first term corresponds to the contribution from modes close to
$k=0$ whereas the second term is due to the peaks at $k=\pi, -\pi$. For the 
case $J_3>0.5$, the contribution from peaks at $k=0, \pi$ and $-\pi$ happens
to be the same as
Eq.~10(a) whereas the contribution $n_1$ from the modes close to $k=k_0$ 
is also equal to 10(a) in the following way:
\setcounter{equation}{10}
\ba
n_1&=&\frac{1}{\pi}\int_0^{\pi}\exp^{[-\pi\tau
\{(\cos k_0-2J_3\cos 2k_0)(k-k_0)\}^2]}\nonumber\\
&=&\frac{1}{2\pi\sqrt{\tau}}[\frac{1}{2J_3+1}+\frac{1}{2J_3-1}].
\ea
The density of kinks 
monotonically increases with increasing $J_3$ provided $J_3 < 0.5$ because of 
the decrease in the off-diagonal term making the probability of excitations
higher. On the other hand, 
for $J_3 >0.5$, the off-diagonal term monotonically increases with with 
increasing $J_3$ resulting to an overall decrease in the density of kinks, 
see figure~3. 
These results can also be seen from the approximate analytical expression
of the kink density given in Eq.~(10 a and b) for both the cases. 

We shall now focus our attention to the case $J_3=0.5$. In the process of
the transverse quenching, the system crosses the multicritical point 
at $h=-0.5, J_3 = 0.5$ as shown in the Fig.~1 and a special power-law behavior
of the kink density is  observed at these parameter values.  
The transition probability
$p_k$ maximizes at $k=0$ as shown above. The argument of the exponential in
$p_k$ is expanded about k=0 at 
$J_3=0.5$, leading to a form $p_k=\exp[-\pi\tau k^6/4]$. 
The contribution to the the kink-density scales as $1/\tau^{1/6}$ which 
can be obtained by simply integrating
this $p_k$, see figure 3. This relatively slow decay of density is a special 
characterisitc of quenching through a multicritical point.
A similar behavior is also
seen in the anisotropic quenching of the transverse XY model \ct{victor07}.  

\begin{figure}[h]
\ig[height=1.9in, width=3.4in]{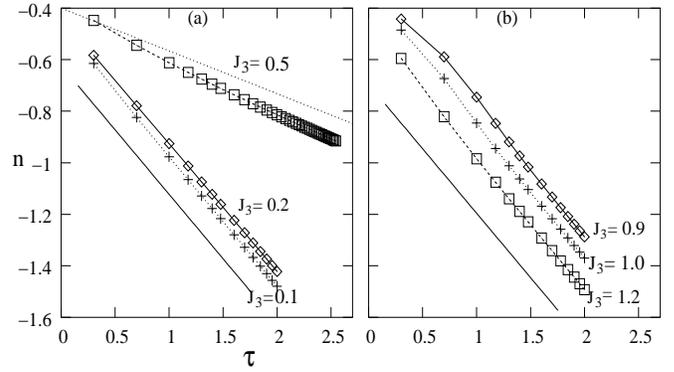}
\caption{The variation of kink density with $\tau$ for different $J_3~(<0.5)$ 
is shown in Fig.~3a. Kink density increases with $J_3$. On the
 other hand for $J_3>0.5$, this variation decreases with $J_3$ as shown in 
Fig.~3b. The thick  line has  slope  $-0.5$ and the slope of the  dotted line
is  $-1/6$, a behavior observed at $J_3=0.5$}
\end{figure}

One can also study  the effect of the anisotropic quenching which 
involves quenching of the nearest neighbor Ising interaction term $J^x(t)
(\sim t/\tau)$ from 
$-\infty$ to 
$+\infty$ instead of the transverse field with the three spin interaction term 
set to unity. 
At $t\to -\infty$, the ground state of the 
system is antiferromagnetic along $x$. 
The probability of the non-adiabatic transition
is similarly given by
\be
p_k = \exp[-\pi\tau\{(h+1)\sin k\}^{2}].
\ee
It is interesting to note that for 
$h=-1$, $p_k$ is unity for all values of $k$.
The density of kinks for the anisotropic quenching is given as
\be
n= \frac {1}{2\pi} \int dk  p_k = \exp((-\pi\tau(h+1)\sin k)^{2})
\ee
with an 
approximate analytical form given as
\be
n = \frac 1 {\pi (h+1) \sqrt{\tau}}
\ee
which shows that the density of kinks decreases monotonically with $h$. This can
be attributed to an increase in the off-diagonal term of the Hamiltonian.

\section{Connection  to the transverse quenching of the Models with next
nearest neighbor interactions}

We shall now use the results of the previous section to
study the transverse quenching of a quantum Ising model with uniform 
ferromagnetic
nearest neighbor interaction and also an additional NNN interaction which  
is either 
antiferromagnetic or ferromagnetic. The model with NNN antiferromagnetic
interaction has 
regular frustrations and is popularly known as  Axial Next Nearest Neighbor
Ising (ANNNI) model \ct{villain} in a transverse field. 
We shall show below that within a mean field approximation, 
the three spin model has a close resemblance to the one-dimensional NNN 
interacting TIMs.

The Hamiltonian of the transverse ANNNI model is given by

\be
H=- \frac 1 {2} \{\sum_i^N[h\sigma^z_{i}+J_1\sigma^x_i\sigma^x_{i+1}- J_2\sigma^x_i\sigma^x_{i+2}]\}
\ee
where $J_1,J_2 >0$.
Henceforth, without loss of any generality, we shall set $J_1=1$. 
At $h=0$, the ground state is ferromagnetically ordered for $J_2 <0.5$,
whereas the system shows 
an ``anti-phase" ordering (where two up spins are followed by two
down spins) for $J_2 >0.5$.  The two phases meet at an
infinitely degenerate multi-critical point $J_2 =0.5$ and $h=0$.
The quantum fluctuations introduced by the  transverse field $h$
competes with the ferromagnetic (or the antiphase) order and eventually 
the system undergoes a quantum
phase transition to a paramagnetic phase at a critical value of the
transverse field given by $h_c$ which is a function of the NNN interaction
$J_2$. 
One-dimensional
quantum ANNNI model shows a rich phase diagram 
which is not fully 
understood till date \ct{dutta96, villain,dutta03}.

When mapped to the corresponding fermionic
Hamiltonian via a JW transformation, the NNN interaction 
leads to a four-fermion term in the fermionic version of the Hamiltonian.
 In the limit $J_2 \to 0$, this
term vanishes so that the model is exactly solvable in terms of non-interacting
fermions. For non-zero $J_2$,
the  fermionic Hamiltonian is  written as
\ba
H=- \frac 1 {2} \{[\sum_ih(2c_i^{\dagger}c_i-1)+(c_i^{\dagger}-c_i)
(c_{i+1}^{\dagger}+c_{i+1})\nonumber\\
-J_2(c_i^{\dagger}-c_i)(1-2c_{i+1}^{\dagger}c_{i+1})
(c_{i+2}^{\dagger}+c_{i+2})]\}.
\ea
The occurrence of the four-fermion term renders the model analytically 
intractable though an approximate analytical solution is possible at
least in the limit of small $J_2$. Deep in the paramagnetic phase
all the spins are oriented in the direction of the transverse field 
so that $<\sigma_i^z>=1$  or in the fermionic language
$1 -2 c_{i+1}^{\dagger}c_{i+1}=-1$. We shall approximate 
$<\sigma_i^z> =1$ for all positive values of $h$ including $h\sim 0$.
This approximation, though crude, transforms the four fermion 
term 
$(c_i^{\dagger}-c_i)(1-2c_{i+1}^{\dagger}c_{i+1})(c_{i+2}^{\dagger}+c_{i+2})$
to a quadratic form.
The Hamiltonian becomes exactly solvable but 
the rich phase diagram of the model is not captured in this approximation
\ct{sen89,sen91}. 
Within this approximation, we shall explore the role of
small NNN antiferromagnetic interaction  on the density of kinks produced 
during the transverse quenching. As described below, this approximation at least
shows a decrease of critical field $h_c$ with $J_2$ for $J_2<0.5$.

The mean field Hamiltonian in the momentum space is
\ba
H=-[\sum_{k>0} (h+{\rm cos}k+J_2{\rm cos}2k)(c_k^{\dagger}c_k+c_{-k}^{\dagger}
c_{-k})\nonumber\\
+{\rm i}({\rm sin}k+J_2{\rm sin}2k)(c_k^{\dagger}c_{-k}^{\dagger}+c_kc_{-k})]
\ea
Comparing Eq.~(17) with Eq.~(3), one finds that the transverse ANNNI chain
Hamiltonian in the mean field approximation is identical to the three spin
Hamiltonian if the antiferromagnetic interaction $J_2$ of the former
is replaced by the negative of the three spin interaction term ($J_3$) 
in the latter. 
Using the results of the previous section, the phase diagram of the mean field 
ANNNI model can be found out for $h>0$(see Fig.~4).
The phase boundary between the ferromagnetic phase and the paramagnetic
is given by $h=1-J_2$ 
with an ordering wave vector $\pi$ (this corresponds to the Ising
transition at $h= J_3 +1$ of Fig.~1). For $J_3 >0.5$, i.e., the transition
between the antiphase and the paramagnetic phase, is given by the corresponding
anisotropic transition of three spin model with the phase boundary given
by the equation $h=J_2$ and the ordering wave vector has an incommensurate
value as given in  Eq.~5.
\begin{figure}
\ig[height=2.2in]{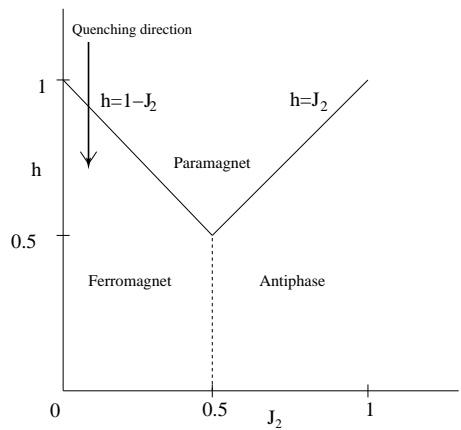}
\caption{Mean field phase diagram of the ANNNI model in the $h-J_2$ plane. We
study the quenching dynamics across the phase boundary close to $J_2 \to 0$.}
\end{figure}

The approximation $1-2c_i^{\dagger}c_i=-1$ is valid for positive $h$
only, we choose a quenching scheme where the transverse field 
has a functional dependence  $h(t)\sim-t/\tau$ with $t$ going from $-\infty$
to $0$  so that h(t) remains positive for
 the entire  quenching period and vanishes at the end of the quenching. 
Therefore the system does not cross the Ising critical line $-h=J_2 +1$.
 
In the final state at $t=0$, all the spins are expected to orient in 
the $x$-direction
with a ferromagnetic order.  
The density of oppositely oriented spins at $t \to 0$ is related to $J_2$ as
\be
n=\frac{1}{2\pi(1-2J_2)\sqrt \tau}
\ee 
which shows that the density of kinks increases monotonically with $J_2$.

It should be noted  that if we follow a  quenching scheme in which
 the transverse field is changed from  $-\infty$ 
to zero, we must
approximate the term $1 -2 c_{i+1}^{\dagger}c_{i+1}$ with $+1$ rather than
$-1$ for above calculations to be viable. In the process of dynamics,
the system crosses the quantum critical line $-h = J_2 +1$ with the modes close 
to $k=0$ getting excited. This approach as well leads to  identical result for
 kink density (as given
in Eq.~18).
Therefore, the presence of a  small antiferromagnetic NNN 
interaction adds to the kink-production in comparison to the ferromagnetic
transverse Ising model ($J_2=0$) with the same quenching scheme.

One can also study, in the similar spirit, a model with a small
 ferromagnetic NNN
interaction $J_{FM}$. We use the same mean field approximation for 
$h \geq 0$ so that
this model is identical to the three spin model with $J_3=J_{FM}$.  A similar 
calculation leads once again to a $1/\sqrt{\tau}$ fall 
of the density of kinks given by
\be
n=\frac{1}{2\pi(1+2J_{FM})\sqrt \tau}.
\ee

\noindent This is expected because the NNN ferromagnetic 
interaction enhances the 
strength of the ferromagnetic ordering
and hence the probability of excitations or density of kinks is lowered.

\section {Conclusions}

In this paper, we have studied the effect of various additional interactions
on the dynamics of the transverse Ising model when swept across the 
quantum critical
lines.
The defect density scale  with the timescale 
 $\tau$ as
$\tau^{-1/2}$, like in transverse Ising case, with a prefactor which
varies from
model to model. 
The first of the variants includes a three spin interaction
with strength $J_3$. Here, the phase 
diagram indicates the existence of an anisotropic phase transition
at an incommensurate value of wave vector in addition to the normal
Ising transition for $J_3>0.5$. Interestingly, we observe that the 
density of kinks 
increases monotonically
with $J_3$ for $J_3<0.5$ whereas decreases for $J_3>0.5$. On the other hand, 
at $J_3=0.5$, the contribution to the kink density scales as $\tau^{-1/6}$
due to the existence of a multicritical point at $J_3=0.5$. 
The other set of Hamiltonians include a ferromagnetic or an antiferromagnetic
next nearest neighbor interactions. The presence of the four fermion term
makes such a Hamiltonian analytically intractable. 
We have used a mean field approximation
to reduce the four fermion term in the fermionized representation 
to a quadratic term. The quenching scheme is chosen carefully
so that the  regions where the approximation is not valid
are avoided in the process of dynamics.
Using the similarity between the  fermionized next nearest neighbor 
interacting Hamiltonians under the
mean field approximation, and  the three spin interacting model, the
density of kink in the final state is estimated.  It is 
observed that the ferromagnetic next nearest neighbor interactions reduces the 
density of kinks produced as opposed to the case of antiferromagnetic 
next nearest neighbor interaction because such a ferromagnetic interaction 
enhances the ferro-ordering discouraging the production of kinks. On the
other hand, frustration leads an enhanced non-adiabatic transitions. We should
mention in conclusion that it is in principle possible to construct a better
mean field theory for the ANNNI model \ct{sen91}, however,  no
qualitative change in the dynamical behaviour in the region $J_2 \to 0$ is
expected.  

We conclude with the comment that the models studied in the 
present work are integrable (at least in the mean field limit) which leads
to a $1/\sqrt{\tau}$ scaling behavior of the defect density. However, in 
a random or a non-integrable system
such a behavior need not be expected \ct{caneva07}. 
The quenching and annealing dynamics of several non-integrable systems
along with the dependence of the defect density on the integrability of 
the model
are yet to be completely understood.
We have also observed a much slower decay
of the form $1/\tau^{1/6}$ when quenched through the
multicritical point of the three spin model 
as in the anisotropic quenching of the transverse XY chain
\ct{victor07}.

\begin{center}
{\bf Acknowledgments}
\end{center}

We acknowledge Victor Mukherjee and Diptiman Sen for collaboration in related
works.

\end{document}